\documentclass[12pt, twoside]{article}
\usepackage{a4wide}
\usepackage{epsfig}

\newsavebox{\uuunit}
\sbox{\uuunit}
    {\setlength{\unitlength}{0.825em}
     \begin{picture}(0.6,0.7)
        \thinlines
        \put(0,0){\line(1,0){0.5}}
        \put(0.15,0){\line(0,1){0.7}}
        \put(0.35,0){\line(0,1){0.8}}
       \multiput(0.3,0.8)(-0.04,-0.02){12}{\rule{0.5pt}{0.5pt}}
     \end {picture}}
\newcommand {\unity}{\mathord{\!\usebox{\uuunit}}}
\begin{document}

\begin{flushright}
hep-th/0309195
\end{flushright}
\vskip 2 cm
\begin{center}
{\Large {\bf Spontaneous localization of bulk fields: the six-dimensional case 
}
}
\\[0pt]

\bigskip
\bigskip {
{\bf Hyun Min\ Lee\footnote{
{{ {\ {\ {\ E-mail: minlee@th.physik.uni-bonn.de}}}}}}},
{\bf Hans Peter\ Nilles\footnote{
{{ {\ {\ {\ E-mail: nilles@th.physik.uni-bonn.de}}}}}}} and
{\bf Max\ Zucker\footnote{
{{ {\ {\ {\ E-mail: zucker@th.physik.uni-bonn.de}}}}}}}
\bigskip }\\[0pt]
\vspace{0.23cm}
{\it Physikalisches Institut der Universit\"at Bonn,} \\
{\it Nussallee 12, 53115 Bonn, Germany.}\\

\bigskip
\vspace{3.4cm} Abstract
\end{center}
We study $N=2$ supersymmetric gauge theories with $d=6$ bulk
and $d=4$ brane fields charged under a U(1) gauge symmetry.
Radiatively induced Fayet-Iliopoulos terms lead to an instability
of the bulk fields. We compute the profile of the bulk zero
modes and observe the phenomenon of spontaneous localization towards 
the position 
of the branes. While this mechanism is quite similar
to the $d=5$ case, the mass spectrum of the excited 
Kaluza-Klein modes shows a crucial difference.

\newpage

 
\section{Introduction}

Recently there has been a revived interest 
in higher dimensional field theories. A
particular fascinating picture is the so-called 
brane world scenario, where different
fields might be confined to space-times (branes) 
of different dimensionality. Such a
picture is expected to be ultimately part of a 
low-energy description of some more
fundamental theory as e.g. string or M-theory. 
In fact, prototypes of this picture can
be found in string theory orbifold compactification \cite{Dixon:jw}
as well as the Horava-Witten
heterotic $E_8 \times E_8$ - M-theory \cite{Horava:1995qa}. In the latter 
case gravitational fields propagate in the
full
$d=11$ space-time dimensions while gauge (and matter) 
fields are confined to the
$d=10$ boundaries of the $d=11$ interval. Orbifold 
compactifications of string theory
typically contain untwisted and (various) twisted 
sectors, where fields propagate in
various sub-spaces confined to fixed points (or surfaces) 
in the compactified
dimensions.

In the attempt to construct realistic models for
 particle physics one therefore
generically faces the situation where e.g. quarks, 
leptons or Higgs bosons originate
from fields of different dimensionalities \cite{Ibanez:1987pj,Font:1988tp}. 
Such a situation is in fact quite
interesting
from the phenomenological point of view as the ``overlap'' 
of the wave function of the
fields in extra dimensions determines various coupling 
constants in the low energy
effective field theory. Mass relations between quarks 
and leptons might therefore
originate from such a higher dimensional mechanism.

Unfortunately, a detailed discussion of these issues 
is quite difficult. While the
set-up
is usually simple when one discusses the picture at tree level, 
complications arise in the
quantum theory. The tree level results, however,  are modified 
substantially even in supersymmetric theories.
In fact, it has been pointed out that 
instabilities appear in many
models with U(1) gauge groups due to the presence of 
radiatively induced 
Fayet-Iliopoulos (FI)-terms \cite{Ghilencea:2001bw}. 
This is particularly relevant 
for phenomenological
applications, as the standard model of particle physics 
contains the U(1) group of
hypercharge.

We are thus interested in a set-up where we have 
simultaneous existence of brane
fields of lower and bulk fields of higher dimensionality. 
Not much is known about the
actual profiles of bulk fields in the presence of brane 
fields. Work up to now has
concentrated on the situation of co-dimension one, i.e. 
$d=5$ bulk fields and $d=4$
brane fields \cite{GrootNibbelink:2002wv}. 
There it was shown that the wave function 
of the bulk zero mode was in
general instable in the presence of FI-terms and that 
this instability could lead to a
spontaneous localization of bulk fields at one of the 
lower dimensional branes. In
this dimensional transmutation \cite{GrootNibbelink:2002qp}, 
the bulk field became 
a brane field and all excited
Kaluza-Klein excitations became heavy and decoupled. 

It is not clear yet, how such a mechanism could be 
understood in the framework of
string theory. The local structure of tadpoles and 
anomalies has been studied in the
framework of heterotic string theory
\cite{Gmeiner:2002es,GrootNibbelink:2003gb,Nibbelink:2003rc}, 
but the explicit 
profile of wave functions of the
bulk field has not been determined yet as the situation is 
far more complicated than in the
co-dimension one case. The latter might also not be 
relevant in all cases, as e. g. in heterotic string
compactification extra dimensions naturally appear in 
complex pairs. Fields typically
live in even dimensions, $d=10$ for the untwisted fields 
and $d=4$ and/or $d=6$ for the
twisted fields. Therefore it would be appropriate to study 
the case of even co-dimension.

In the present paper we consider the simplest case: 
$d=4$ brane fields and $d=6$ bulk
fields. This discussion should be understood as a building
block to eventually go
towards the full $d=10$ theory. It illustrates the specific 
features of the theory with
compactified complex dimension in a framework where 
explicit calculations can be
performed (despite the fact of the missing off-shell 
formulation of $d=6$ supersymmetry for hypermultiplets). 
Effective potential, ground state 
wave function and mass spectrum
of the bulk fields can be determined. We show that 
the presence of localized FI-tadpoles 
leads to a localization phenomenon similar 
to the $d=5$ (co-dimension one)
case. The mass spectrum of the Kaluza-Klein modes, 
however, reveals a profound
difference to the co-dimension one situation. The 
$d=6$ bulk field retains its six
dimensional nature and the dimensional transmutation 
of the bulk field to a brane field seems to be a
particular property of the co-dimension one case. 
This might have important
consequences for the discussion of localized anomalies.

The paper is organized as follows. In section 2 we 
give the set-up of supersymmetry
of the $T^2/ {\bf Z}_2$ orbifold. Section 3 presents the 
FI-tadpoles in the $d=6$ case. This is
followed in section 4 with the derivation of 
the effective potential, its minimization
and the calculation of mass spectrum and 
zero-mode wave function. Section 5
discusses possible embeddings in existing string models 
and the question of anomaly
cancellation by (variants of) the Green-Schwarz mechanism. 
In section 6 we give
some concluding remarks.


\section{Supersymmetry on $T^2/{\bf Z}_2$}
In this section we describe in detail our model: a six-dimensional $N=2$ super Yang-Mills multiplet 
with gauge group $U(1)$, coupled to several hypermultiplets. 
This theory is compactified on the orbifold $T^2/{\bf Z}_2$. In addition, 
we have chiral multiplets, living on the fixed points of the orbifold. They are charged under the $U(1)$.
\subsection{The super Yang-Mills multiplet}
The $N=2$ super Yang-Mills multiplet in six dimensions contains 
as propagating fields the gauge field $A_M$ and the gaugino $\Omega$, 
which is a right handed symplectic Majorana-Weyl fermion,
 satisfying the chirality condition (our conventions are collected in appendix \ref{appa})
\begin{equation}
\Gamma^7\Omega=\Omega.\label{ch}
\end{equation}
The gaugino is an eight-component object and a doublet under\footnote{The 
automorphism group of the $N=2$ supersymmetry algebra 
in six dimensions is $SU(2)_{\cal R}$, 
so that all fields belonging to $N=2$ supermultiplets live in representations of this group.}
$SU(2)_{\cal R}$.
Beside these two propagating fields, there is an auxiliary field $\vec{D}$ which is a triplet under the $SU(2)_{\cal R}$. The action for abelian gauge group is
\begin{equation}
{\cal L}=-\frac{1}{4}F_{MN}F^{MN}+i\bar{\Omega}\Gamma^M\partial_M\Omega+\frac{1}{2}\vec{D}^2.\label{lag1}
\end{equation}
The Lagrangian is invariant under (rigid) supersymmetry:
\begin{eqnarray}
\delta A_M & = &i\bar{\varepsilon}\Gamma_M\Omega\\
\delta \Omega & = & \frac{1}{4}\Gamma^{MN}\varepsilon F_{MN}-\frac{i}{2}\vec{\tau}\varepsilon\vec{D}\\
\delta\vec{D} & = & \bar{\varepsilon}\vec{\tau}\Gamma^M\partial_M\Omega.
\end{eqnarray}
The supersymmetry parameter $\varepsilon$ is also a right handed symplectic Majorana-Weyl fermion, satisfying a similar relation as the gaugino, eq. (\ref{ch}).

The $\Gamma$-matrices are $8\times 8$ matrices.
By using the chiral representation given in (\ref{rep1})-(\ref{rep4}), the chirality constraints can be solved, i.e. we can work in a four-dimensional Weyl representation. Defining for an arbitrary spinor
\begin{eqnarray}
\psi={\psi_L \choose \psi_R}
\end{eqnarray}
and using the $4\times 4$ matrices
\begin{eqnarray}
\gamma^A\equiv (\gamma^a,\gamma^5,-\unity_4)\qquad \mbox{and}\qquad \bar{\gamma}^A\equiv (\gamma^a,\gamma^5,\unity_4),
\end{eqnarray}
the action (\ref{lag1}) can be rewritten as
\begin{equation}
{\cal L}=-\frac{1}{4}F_{MN}F^{MN}+i\bar{\Omega}_R\gamma^M\partial_M\Omega_R+\frac{1}{2}\vec{D}^2.\label{symact}
\end{equation}
The transformation laws become
\begin{eqnarray}
\delta A_M & = &i\bar{\varepsilon}_R\gamma_M\Omega_R\\
\delta \Omega_R & = & \frac{1}{4}\gamma^{MN}\varepsilon_R F_{MN}-\frac{i}{2}\vec{\tau}\varepsilon_R\vec{D}\\
\delta\vec{D} & = & \bar{\varepsilon}_R\vec{\tau}\gamma^M\partial_M\Omega_R.
\end{eqnarray}
The symplectic Majorana condition for the right-handed Dirac gaugino becomes
\begin{eqnarray}
{\bar\Omega}_{Ri}=\varepsilon_{ij}(\Omega^j_R)^T{\cal C},
\end{eqnarray}
where ${\cal C}$ is the five-dimensional charge conjugation as given 
in the appendix. 
Then, we can solve this condition by writing the gaugino in terms of one Dirac spinor $\chi$ as
\begin{eqnarray}
\Omega_{R}^i=\left(\begin{array}{l} \,\,\,\,\chi \\ {\cal C}{\bar\chi}^T
\end{array}\right).
\end{eqnarray}
Similarly, we can also solve the chirality condition for $\varepsilon_R$.
By dimensionally reducing the super Yang-Mills multiplet to five dimensions and comparing with the known results \cite{Zucker:2003qv, Zucker:1999fn}, 
one can check the consistency of the present theory.

\subsection{The hypermultiplet}
The other multiplet needed is the hypermultiplet. 
For this multiplet there is no off-shell formulation possible, 
since this would require the fields to be charged under central charge 
transformations. 
However, in contrast to five dimensions \cite{Zucker:2003qv}, 
in six-dimensional $N=2$ supersymmetry there is no central charge, 
which is due to the fact that $d=6$ is the highest dimension 
where $N=2$ supersymmetry can exist. 
Fortunately, the non-existence of an off-shell formulation poses no problem 
for our purposes.

The $r$ hypermultiplets contain $4r$ real scalars $A_i^\alpha$ 
and the hyperino $\zeta^{\alpha}$, where the gauge index has to run over 
an even number of values, $\alpha,\beta = 1,\ldots ,2r$. 
The generators of the representation are anti-hermitian 
$t^{\alpha}{}_\beta=-t_\beta{}^\alpha$, 
where $t_\beta{}^\alpha=(t^\beta{}_\alpha)^*$.
 The bosons transform in the $\bf 2$ of $SU(2)_{\cal R}$, whereas the fermions are singlets. The scalars $A_i^\alpha$ satisfy a reality condition
\begin{equation}
A^i_\alpha\equiv A_i^{\alpha *}=\varepsilon^{ij}\rho_{\alpha\beta}A_j^\beta\label{realcon}
\end{equation}
where $\rho$ can be choosen to be $\rho=\unity\otimes\varepsilon$, as shown in \cite{deWit:1984px}. 
 Consistency requires a reality constraint on the generators of the gauge group,
\begin{eqnarray}
t^\alpha{}_\beta=-\rho^{\alpha\gamma}t_\gamma{}^\delta\rho_{\delta\beta}.\
\end{eqnarray}

The chirality of the hyperino is opposite to the one of the gaugino, i.e.
\begin{eqnarray}
\Gamma^7\zeta^{\alpha}=- \zeta^{\alpha}.
\end{eqnarray}
The supersymmetry transformation laws together with the reality constraint (\ref{realcon}) induce a reality constraint for the hyperino,
\begin{equation}
\bar{\zeta}_\alpha = -\rho_{\alpha\beta}{\zeta^{\beta}}^TC,\label{real22}
\end{equation}
where the Dirac conjugate is defined in the standard way (cf. eq. (\ref{SMB})).

The supersymmetry transformation laws we find are
\begin{eqnarray}
\delta A_i^\alpha & = &i\bar{\varepsilon}_i\zeta^\alpha\\
\delta \zeta^\alpha & = & -\Gamma^{A}\varepsilon^i {\cal D}_A A_i^\alpha,
\end{eqnarray}
with covariant derivative
\begin{eqnarray}
{\cal D}_M A_i^\alpha=\partial_MA_i^\alpha-g(A_M)^{\alpha}{}_\beta A_i^\beta.
\end{eqnarray}
The supersymmetry algebra closes only up to equations of motion. 

The hypermultiplet Lagrangian is
\begin{eqnarray}
{\cal L}=\frac{1}{2}{\cal D}_MA_i^\alpha{\cal D}^M A^i_\alpha+\frac{i}{2}\bar{\zeta}_\alpha\Gamma^M{\cal D}_M\zeta^\alpha-2ig\bar{\zeta}_\alpha\Omega^{i\alpha}{}_\beta A^\beta_i+\frac{ig}{2}(\vec{\tau})^i{}_j A_i^\alpha A^j_\beta\vec{D}^\beta{}_\alpha.
\end{eqnarray}
We can again use the chiral basis (\ref{rep1})-(\ref{rep4}) which 
we used already for the super Yang-Mills multiplet. 
The Lagrangian then becomes
\begin{equation}
{\cal L}=\frac{1}{2}{\cal D}_MA_i^\alpha{\cal D}^M A^i_\alpha+\frac{i}{2}\bar{\zeta}_{L\alpha}\bar{\gamma}^M{\cal D}_M\zeta^\alpha_L-2ig\bar{\zeta}_{L\alpha}\Omega_R^{i\alpha}{}_{\beta} A^\beta_i+\frac{ig}{2}(\vec{\tau})^i{}_j A_i^\alpha A^j_\beta\vec{D}^\beta{}_\alpha\label{hyperact}
\end{equation}
and the transformation laws read
\begin{eqnarray}
\delta A_i^\alpha & = &i\bar{\varepsilon}_{Ri}\zeta^\alpha_L\\
\delta \zeta^\alpha_L & = & -\gamma^{A}\varepsilon^i_R {\cal D}_A A_i^\alpha.
\end{eqnarray}

We can rewrite the Lagrangian in a better suited way. 
To this end, we introduce a two component field \cite{Ohashi}
\begin{eqnarray}
A_i^{\hat{\alpha}(i)}={A_i^{2\hat{\alpha}-1}\choose A_i^{2\hat{\alpha}}},
\end{eqnarray}
where the hatted index runs over $\hat{\alpha}=1,\ldots ,r$ and the index in brackets denotes the two entries in the object on the r.h.s. The reality constraint (\ref{realcon}) now becomes diagonal,
\begin{eqnarray}
A^i_{\hat{\alpha}(i)}=\delta_{\hat{\alpha}\hat{\beta}}\varepsilon_{(i)(j)}\varepsilon^{ij}A_j^{\hat{\beta}(j)}.
\end{eqnarray}
This equation can easily be solved by introducing complex fields $\phi_\pm$
\begin{equation}
A_i^{\hat{\alpha}(i)}\equiv
\left(\begin{array}{cc}
A_1^{\hat{\alpha}(1)}&A_2^{\hat{\alpha}(1)}\\
A_1^{\hat{\alpha}(2)}&A_2^{\hat{\alpha}(2)}\\
\end{array}\right)=
\left(\begin{array}{cc}
\phi_-^{*\hat{\alpha}}&\phi_+^{\hat{\alpha}}\\
-\phi_+^{*\hat{\alpha}}&\phi_-^{\hat{\alpha}}\\
\end{array}\right),\label{gamm}
\end{equation}
which makes the underlying quaternionic structure more explicit 
(cf. \cite{deWit:1984px, Ohashi, Bergshoeff:1985mz}).
The meaning of the index $\pm$ will become clear when we orbifold 
the hypermultiplet. 

Likewise, the reality constraint (\ref{real22}) for the hyperino 
with a two component field $\zeta^{{\hat\alpha}(i)}_L$ becomes
\begin{eqnarray}
\bar{\zeta}_{L{\hat\alpha}(i)}
=\delta_{{\hat\alpha}{\hat\beta}}\varepsilon_{(i)(j)}
(\zeta^{{\hat\beta}(j)}_L)^T{\cal C}.
\end{eqnarray}

Then, this equation is solved by introducing one Dirac spinor $\psi$ as
\begin{eqnarray}
\zeta^{{\hat\alpha}(i)}_L=\left(\begin{array}{l}\,\,\,\,\,\,\,\,\psi 
\\ {\cal C}{\bar\psi}^T
\end{array}\right).
\end{eqnarray}

Using the $\phi_\pm$ and $\psi$ fields, 
the relevant hypermultiplet Lagrangian from eq.~(\ref{hyperact})
becomes (suppressing $\hat{\alpha}$-indices)
\begin{eqnarray}
{\cal L}=\sum_\pm\left({\cal D}_M\phi_\pm^\dagger{\cal D}^M\phi_\pm
\mp g\phi_\pm^\dagger q\phi_\pm D_3\right)+i{\bar\psi}\bar{\gamma}^M{\cal D}_M\psi+\ldots.
\end{eqnarray}
Here we have chosen the gauge group to be $U(1)$ and the generators 
to be $t=iQ=-iq\otimes \tau^3$, with diagonal charge matrices $Q$ and $q$.
The covariant derivatives are
\begin{eqnarray}
{\cal D}_M\phi_\pm=\partial_M\phi_\pm\pm igq\phi_\pm A_M.
\end{eqnarray}

\subsection{Orbifolding}
We now consider our theory given by the sum of the actions (\ref{symact}) 
and (\ref{hyperact}) on the orbifold $T^2/{\bf Z}_2$. 
The coordinates which form the torus are $x^5$ and $x^6$ with radii $R_5$ 
and $R_6$, respectively.
The ${\bf Z}_2$ acts like
\begin{eqnarray}
{\bf Z}_2:\qquad (x^5,x^6)\to (-x^5,-x^6).
\end{eqnarray}
So the $T^2/{\bf Z}_2$ orbifold has four fixed points,
\begin{equation}
(x^5,x^6)=\qquad (0,0),\qquad (\pi R_5, 0),\qquad (0,\pi R_6),\qquad\mbox{and}\qquad (\pi R_5,\pi R_6).\label{fixpoints}
\end{equation}

A boson is either even (${\cal P}=+1$) or odd (${\cal P}=-1$) under the ${\bf Z}_2$. The parities $\cal P$  of the bosons belonging to the super Yang Mills multiplet are collected in the table.
 \begin{table}
\begin{center}
\begin{tabular}{|c|c|c|c|c|c|c|}
\hline\hline
Field & $A_m$ & $A_5$ & $A_6$ & $D_1$ & $D_2$ & $D_3$\\
\hline
Parity $\cal P$&$+1$&$-1$&$-1$&$-1$&$-1$&$+1$\\
\hline
\end{tabular}
\end{center}
\caption{Parities of the super Yang-Mills multiplet.}\label{table3022}
\end{table}

The gaugino transforms like
\begin{eqnarray}
\Omega(x^m,x^5,x^6)\to -i\tau^3\Gamma_5\Gamma_6\Omega(x^m,-x^5,-x^6)
\end{eqnarray}
which becomes in the chiral representation we used repeatedly 
\begin{eqnarray}
\Omega(x^m,x^5,x^6)_R\to i\tau^3\gamma^5\Omega(x^m,-x^5,-x^6)_R.
\end{eqnarray}
At the fixed points, the even fields form a four-dimensional 
$N=1$ super Yang-Mills multiplet with field content
\begin{eqnarray}
(A_m, \Omega_R, -D_3+F_{56}),
\end{eqnarray}
i.e. the four-dimensional auxiliary field is not $D_3$ 
as one might have naively expected but $-D_3+F_{56}$. 
We can compare this to the five-dimensional result 
(see, e.g. \cite{Mirabelli:1997aj, GrootNibbelink:2002wv, GrootNibbelink:2002qp}). 
There the four-dimensional auxiliary field was given by $-D_3+\partial_5\Phi$, 
where $\Phi$ is to be identified with $A_6$. 
So $-D_3+F_{56}$ is the gauge covariant generalization 
of $-D_3+\partial_5 \Phi$.
 
The orbifolding of the hypermultiplet is most easily written down 
using the $\phi_\pm$-fields. One finds
\begin{eqnarray}
{\bf Z}_2:\qquad \phi_\pm^{\hat{\alpha}}\to \pm \phi_\pm^{\hat{\alpha}}
\end{eqnarray}
and for the hyperino
\begin{eqnarray}
\zeta^{\hat{\alpha}(i)}_L\to i(\tau^3)^{(i)}{}_{(j)}\gamma^5 \zeta^{\hat{\alpha}(j)}_L.
\end{eqnarray}

\subsection{Chiral multiplets at the fixed points}
As stated above, the super Yang-Mills multiplet forms a four-dimensional super Yang-Mills multiplet at the fixed points. In the spirit of Mirabelli and Peskin \cite{Mirabelli:1997aj}, we can couple chiral multiplets which live 
at the fixed points to the super Yang-Mills multiplet. 
Our formalism is chosen to agree at the fixed points 
with the one in \cite{GrootNibbelink:2002wv, GrootNibbelink:2002qp}, 
so our Lagrangian is the same as given there. 
For completeness, we should give it here:
\begin{eqnarray}
{\cal L} & = & \sum_{I=1}^4\delta(x^5-x^5_I)\delta(x^6-x^6_I)\Big[{\cal D}_m\phi_I^{\dagger}{\cal D}^m\phi_I \nonumber \\
& + & i\bar{\psi}_{I\ell}\gamma^m{\cal D}_m\psi_{I\ell}
+ f^\dagger_If_I+g\phi^\dagger_Iq_I\phi_I(-D_3+F_{56})+\cdots\Big].
\end{eqnarray}
$(x^5_I, x^6_I)$ label the fixed points as given in (\ref{fixpoints}). 
$q_I$ are the charge matrices at the fixed points. 
More details on the boundary Lagrangian may be found 
in \cite{GrootNibbelink:2002wv, GrootNibbelink:2002qp}.

\section{Tadpoles}
We want to calculate the tadpole diagrams which generate Fayet-Iliopoulos terms at the branes. The procedure is exactly as in \cite{GrootNibbelink:2002wv, GrootNibbelink:2002qp} and the result can be obtained accordingly. 
We shall therefore not present the details of the calculation here.
 
The tadpole diagrams contain the hyperons $\phi_\pm$ and the hyperinos $\psi$
as well as the charged brane scalars $\phi_I$. 

As discussed above, the $D$ field belonging to the four-dimensional super 
Yang-Mills multiplet is given by $D=-D_3+F_{56}$.
So the form of our Fayet-Iliopoulos term is 
\begin{eqnarray}
{\cal L}_{FI}=\xi(-D_3+F_{56}).
\end{eqnarray}
There are two types of contributions to $\xi$, 
one from the bulk, the other one from the branes,
\begin{eqnarray}
\xi=\xi_{bulk}+\xi_{branes}
\end{eqnarray}
with
\begin{eqnarray}
\xi_{bulk}=\frac{1}{4}g~ {\rm tr}(q)\left(\frac{\Lambda^2}{16\pi^2}+\frac{1}{4}\frac{\ln \Lambda^2}{16\pi^2}(\partial_5^2+\partial_6^2)\right)\sum_I\delta(x^5-x^5_I)\delta(x^6-x^6_I)
\end{eqnarray}
and
\begin{eqnarray}
\xi_{brane}=g\frac{\Lambda^2}{16\pi^2}\sum_I {\rm tr}(q_I)\delta(x^5-x^5_I)\delta(x^6-x^6_I).
\end{eqnarray}
This result for $\xi$ is, of course, similar to the one obtained 
in the five-dimensional $S^1/{\bf Z}_2$ model \cite{GrootNibbelink:2002wv,GrootNibbelink:2002qp}: 
The main difference is an additional factor of $1/2$ in $\xi_{bulk}$ 
due to the double number of fixed points in our model. 
The only other difference $\partial^2_5\to\partial_5^2+\partial_6^2$ 
should be clear. 
The brane contributions are unchanged. 

We reshuffle the contributions and write the Fayet-Iliopoulos term as the sum of
quadratically divergent and logarithmically divergent pieces:
\begin{eqnarray}
\xi=\sum_I\left(\xi_I+\xi''(\partial_5^2+\partial_6^2) \right)\delta(x^5-x^5_I)\delta(x^6-x^6_I)
\end{eqnarray}
with
\begin{eqnarray}
\xi_I&=&\frac{1}{16\pi^2}g\Lambda^2\left(\frac{1}{4} {\rm tr}(q)+ {\rm tr} (q_I)\right),\\ 
\xi''&=&\frac{1}{16}\frac{1}{16\pi^2}g\ln \Lambda^2 {\rm tr} (q).
\end{eqnarray}
The distribution of quadratic FI tadpoles $\xi_I$'s 
on the orbifold is shown in Fig.~1.

\begin{figure}[ht]
\begin{center}
\epsfig{file=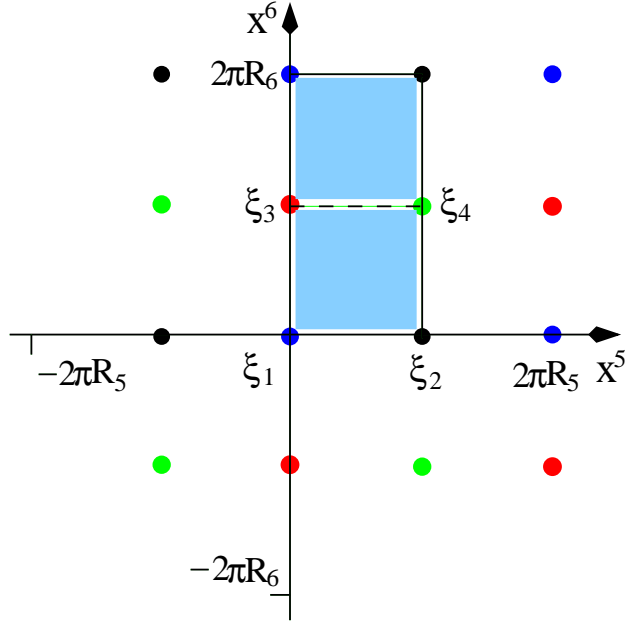,width=8.25cm}
\caption{ \label{orbifold}
The distribution of localized Fayet-Iliopoulos tadpoles 
on a $T^2/{\bf Z}_2$ orbifold. The shaded region is the fundamental region of the orbifold.}
\end{center}
\end{figure}

\section{The effective potential}
It is now easy to write down the effective four-dimensional potential; 
we simply collect all pieces of the action which give rise to the potential 
in the four-dimensional theory:
%
\begin{eqnarray}
V&=&\int dx^5 dx^6\Big[2g^2|\phi^T_+q\phi_-|^2
+\sum_\pm({\cal D}_5\phi_\pm+i{\cal D}_6\phi_\pm)^\dagger({\cal D}_5\phi_\pm+i{\cal D}_6\phi_\pm)\nonumber \\
&+& \frac{1}{2}\Big(F_{56}-g(\phi_+^{\dagger} q\phi_+-\phi_-^{\dagger} q\phi_-)
-\xi-g\sum_I\delta(x^5-x^5_I)\delta(x^6-x^6_I)\phi^\dagger_I q_I\phi_I\Big)^2 \nonumber \\
&-& \frac{1}{2}\Big(D_3-g(\phi_+^{\dagger} q\phi_+-\phi_-^{\dagger} q\phi_-)
-\xi-g\sum_I\delta(x^5-x^5_I)\delta(x^6-x^6_I)\phi^\dagger_I q_I\phi_I\Big)^2\nonumber \\
&-& \frac{1}{2}(D_1+g\phi^T_+ q \phi_-+g\phi^\dagger_- q \phi_+^*)^2
-\frac{1}{2}(D_2+ig\phi^T_+ q \phi_--ig\phi^\dagger_- q \phi_+^*)^2 
\Big]
\end{eqnarray}
A consistency check is to show that unbroken supersymmetry is equivalent to vanishing potential. After elimination of the auxiliary fields, the potential is positive semi-definite.

The conditions for unbroken supersymmetry are
\begin{equation}
D_3=F_{56}=g(\phi_+^{\dagger} q\phi_+-\phi_-^{\dagger} q\phi_-)+\xi+g\sum_I\delta(x^5-x^5_I)\delta(x^6-x^6_I)\phi^\dagger_I q_I\phi_I\label{cond1}
\end{equation}
together with
\begin{eqnarray}
\phi_+^Tq\phi_-=0
\end{eqnarray}
and
\begin{equation}
{\cal D}_5\phi_\pm+i{\cal D}_6\phi_\pm=0.\label{third}
\end{equation}

\subsection{Supersymmetric background solutions}
Now suppose that our ground state does not break the $U(1)$, i.e.\footnote{We assume that all fields are charged under the $U(1)$.}
$\langle \phi_\pm\rangle=\langle \phi_I\rangle=0$.
Then, the supersymmetry condition (\ref{cond1}) becomes 
\begin{equation}
\langle F_{56}\rangle=\xi.\label{back}
\end{equation}
When we integrate this equation over the extra dimensions, Stokes theorem with no boundary gives
\begin{eqnarray}
0=\int^{\pi R_5}_0 dx^5\int^{\pi R_6}_0 dx^6 \langle F_{56}\rangle=\sum_I\xi_I,
\end{eqnarray}
which means
\begin{equation}
{\rm tr}(q)+{\rm tr}(q_1)+{\rm tr}(q_2)+{\rm tr}(q_3)+{\rm tr}(q_4)=0.
\label{sum}
\end{equation}
This consistency condition ensures the absence
of overall mixed gauge-gravitational anomalies. If (\ref{sum}) is
violated, we would expect the U(1) to be broken at a high scale
either spontaneously or through a variant of a Green-Schwarz mechanism.
We shall come back to this question in section 5. 

Now let us introduce complex coordinates
\begin{equation}
z=\frac{1}{R_5}x^5+\frac{1}{R_6}\tau x^6,\label{cc}
\end{equation}
with the torus modulus $\tau=iR_6/R_5$.
By this definition, the periodicities on the torus are
\begin{eqnarray}
z\simeq z+2\pi\simeq z+2\pi\tau.
\end{eqnarray}

A key obervation is the fact that we can solve 
eq.~(\ref{back}) with the following ansatz:
\begin{equation}
\langle A_5\rangle=-\partial_6 W\qquad
\mbox{and}\qquad \langle A_6\rangle=\partial_5 W\label{W},
\end{equation}
where $W$ must be even under ${\bf Z}_2$. Please note that by taking this ansatz we have fixed the gauge implicitly.
Defining
\begin{equation}
W'=2\left(W-\frac{2}{R^2_5}\sum_I\xi''\delta^2(z-z_I)\right),\label{def2}
\end{equation}
eq. (\ref{back}) becomes
\begin{equation}
\partial\bar{\partial}W'=\sum_I\xi_I\delta^2(z-z_I),\label{Poisson}
\end{equation}
which is some sort of Poisson equation.

The solution to eq.~(\ref{Poisson}) can be deduced 
from string theory: $W'$ is the propagator of a bosonic string 
for a toroidal world sheet, see e.g. \cite{Polchinski:rq}. 
The modular invariant and periodic solution to ({\ref{Poisson}) 
on the torus is then
\begin{equation}
W'=\frac{1}{2\pi}\sum_I\xi_I
\left[\ln\left|\vartheta_1\left(\frac{z-z_I}{2\pi}\big|
\tau\right)\right|^2-\frac{1}{2\pi\tau_2}[{\rm Im}(z-z_I)]^2\right]
\label{sol}
\end{equation}
where $\tau_2={\rm Im}\tau=R_6/R_5$.
Note that in order for $W'$ in the above to be a solution 
to (\ref{Poisson}), eq. (\ref{sum}) has to hold. 
We can also check that $W'$ is invariant under ${\bf Z}_2$ 
as required from eqs.~(\ref{W}) and (\ref{def2}).

\subsection{Localization of the bulk zero mode}

Now we are in the position to consider the solution for the zero mode 
in the presence of the background solution.
The equation for the bulk zero mode is the same as eq.~(\ref{third}).
Defining the complex potential $A=A_5-iA_6$ in the complex coordinates 
from eq.~(\ref{cc})
and using the background solution with $A=-(2i/R_5)\partial W$ 
from eq.~(\ref{W}), 
the equation for the zero mode becomes
\begin{eqnarray}
({\bar \partial}-gq{\bar \partial}W)\phi_+=0.
\end{eqnarray}
Thus, from eqs.~(\ref{def2}) and (\ref{sol}), we obtain the exact solution 
for the zero mode as
\begin{eqnarray}
\phi_+&=&f_+(z) e^{gqW} \nonumber \\
&=&f_+(z)\prod_I \left|\vartheta_1\left(\frac{z-z_I}{2\pi}\big|
\tau\right)\right|^{\frac{1}{2\pi}gq\xi_I}\times \nonumber \\
&\times&\exp\left[-\frac{1}{8\pi^2\tau_2}gq\xi_I[{\rm Im}(z-z_I)]^2
+ \frac{gq\xi''}{R^2_5}\delta^2(z-z_I)\right], 
\label{final}
\end{eqnarray}
where $f_+(z)$ is a holomorphic function of $z$. 
Since $f_+(z)$ is an analytic function in the whole complex plane 
where ${\bar\partial}f_+=0$, 
it should be a constant which is determined 
by the normalization condition
\begin{eqnarray}
1=\int_0^{\pi R_5}dx^5\int_0^{\pi R_6}dx^6|\phi_+|^2.
\end{eqnarray}

Let us now discuss the localization of the zero mode. There are three factors
to be discussed from the wave function of the zero mode (\ref{final}): 
the $\vartheta_1$ term, the $e^{({\rm Im})^2}$ 
term and the $e^{\delta^2}$ term.
The second one implies a (de)localization behavior of zero mode 
at $z=z_I$ with $q\xi_I<0(q\xi_I>0)$. From 
the asymptotic limit of the theta function for $z\rightarrow z'$ 
\begin{eqnarray}
\vartheta_1\bigg(\frac{z-z'}{2\pi}|\tau\bigg)\rightarrow 
(\eta(\tau))^3 (z-z'),
\end{eqnarray}
where $\eta(\tau)$ is the Dedekind eta function,
we can see that the wave function of the zero mode becomes divergent 
at the fixed points where $q\xi_I<0$. 
In fact, from eq.~(\ref{sum}), at least one of the $\xi_I$'s should
take a different sign from the others, so it would imply  
the strong localization of the zero mode at up to three fixed points 
at the same time. 
Moreover, the $e^{\delta^2}$ term also seems to give a strong (de)localization
for $q\xi''>0(q\xi''<0)$ as in the five-dimensional case \cite{GrootNibbelink:2002wv}. 
Thus the $\vartheta_1$ term with $q\xi_I<0$ and the $e^{\delta^2}$ term need
to be regularized. 
 
By inserting a simple regularization of the delta 
function in eqs.~(\ref{def2}) and (\ref{Poisson}) with 
\begin{eqnarray}
\delta^2(z-z_I)=\left\{\begin{array}{l}\frac{R^2_5}{\pi\rho^2},\qquad  
|z-z_I|<\rho/R_5, \\
0, \qquad\,\,\,\, |z-z_I|>\rho/R_5,
\end{array}\right.
\end{eqnarray}
and omitting the normalization, the regularized zero mode function 
for $|z-z_I|\ll 1$ is given for a finite $\rho$ with $\rho\ll R_5$ as 
\begin{eqnarray}
\phi_+\simeq\left\{\begin{array}{l} 
\exp\bigg[\frac{gq}{4\pi}\xi_I\ln\big(\frac{\rho^2}{R^2_5}\big)
+\frac{gq\xi''}{\pi\rho^2}\bigg],\qquad |z-z_I| <\rho/R_5, \\
\exp\bigg[\frac{gq}{4\pi}\xi_I\bigg(\ln(|z-z_I|^2)
-\frac{1}{2\pi\tau_2}[{\rm Im}(z-z_I)]^2\bigg)\bigg], 
\,\, \rho/R_5 <|z-z_I|\ll 1.\label{limit}
\end{array}\right.
\end{eqnarray}
To understand the localization of the zero mode explicitely we have to
maintain two regularization scales: the momentum cutoff $\Lambda$ and the
brane thickness $\rho$; both $\rho$ and $1/\Lambda$ are small compared to
$R_5, R_6$. The localization induced by $\xi$ is typically exponential
in $\Lambda$ while the one induced by $\xi''$ is power like. Thus as 
long as $\rho$ is not very small compared to $1/\Lambda$ the effect of the 
logarithmic FI-term will be subleading (naturally one could expect
$\rho$ and  $1/\Lambda$ to be of the same order of magnitude). In
addition, as we shall see in the next section, the logarithmic
FI-term does not effect the mass spectrum of the bulk field at all.

\subsection{Mass spectrum}

Making a Kaluza-Klein reduction to $d=4$, the equation for the massive modes 
with nonzero gauge field background is given by
\begin{eqnarray}
(-{\cal D}_5+i{\cal D}_6)({\cal D}_5+i{\cal D}_6)\phi_\pm=m^2\phi_\pm.
\end{eqnarray}
Then, this equation can be rewritten in terms of the scalar function $W$, 
given by the gauge field solutions in eq.~(\ref{W}), as
\begin{equation}
(\partial\pm gq\partial W)(\bar{\partial}\mp gq\bar\partial W)\phi_\pm=-\frac{1}{4}m^2R^2_5\phi_\pm.\label{eom1}
\end{equation}
By substituting in eq. (\ref{eom1})
\begin{eqnarray}
\phi_\pm = e^{\pm gqW}\tilde{\phi}_\pm,
\end{eqnarray}
we get a simpler form
\begin{eqnarray}
\partial\bar{\partial}\tilde{\phi}_\pm\pm 2gq\partial W\bar{\partial} 
\tilde{\phi}_\pm=-\frac{m^2}{4}R_5^2\tilde{\phi}_\pm.\label{eom2}
\end{eqnarray}
Since there appear derivatives of delta functions in this equation,
we need to regularize the delta function. 
Let us take the regularizing function to be
\begin{eqnarray}
\Delta^2(z-z_I)=\left\{ \begin{array}{cl}
\frac{R^2_5}{\pi \rho^2_I}\bigg(1-\frac{R^2_5|z-z_I|^2}{\rho^2_I}\bigg), 
&\mbox{for }|z-z_I|<\rho_I/R_5, \\
0, &\mbox{for }|z-z_I|>\rho_I/R_5,
\end{array}\right.
\end{eqnarray} 
where $\rho_I$ corresponds to the thickness of the $z=z_I$ brane. 
Here one can easily check that 
$\lim_{\rho_I\rightarrow 0}\int d^2 z'\Delta^2(z-z')h(z',{\bar z}')=h(z,{\bar z})$ for an arbitrary complex function $h$. 
Then, the solution for $W$ given in eq.~(\ref{def2}) becomes
\begin{eqnarray}
W=\frac{1}{2}\int d^2 z'\, G(z-z')\sum_I\xi_I\Delta^2(z'-z_I)
+\frac{2}{R^2_5}\sum_I\xi^{\prime\prime}\Delta^2(z-z_I),
\end{eqnarray}
where $G(z-z')$ is the string propagator on the torus satisfying 
${\bar \partial}\partial G(z-z')=\delta^2(z-z')-1/(8\pi^2\tau_2)$.
In order for $W$ to be periodic on the torus, which is necessary even in view of
the zero mode in eq.~(\ref{final}), 
we need the same regularization of branes, i.e. the same $\rho_I$'s. 
In this case, we again have the zero sum $\sum_I\xi_I=0$. 
Consequently, we get the holomorphic derivative of $W$ as
\begin{eqnarray}
\partial W=\frac{1}{2}\sum_I\xi_I\int d^2 z'\,\partial G(z-z')\Delta^2(z'-z_I)
+\frac{2}{R^2_5}\sum_I\xi^{\prime\prime}\partial\Delta^2(z-z_I).\label{pW}
\end{eqnarray}
Now let us consider the region outside the brane 
where $\partial W$ is holomorphic because ${\bar\partial}(\partial W)=0$. 
Then, for $|z-z_I|>\rho_I/R_5$, 
eq.~(\ref{eom2}) becomes solvable
with a separation of variables
as ${\tilde\phi}_{\pm}=\chi_\pm(z)\,\varphi_{\pm}({\bar z})$ 
\begin{eqnarray}
&&\bigg[\partial(\ln\chi_\pm\pm 2gq\sum_I\xi_I\int d^2 z'\,
f(z-z')\Delta^2(z'-z_I)) \nonumber \\
&\pm& \frac{gq}{8\pi^2\tau_2}\sum_I\xi_I{\bar z}_I\bigg] 
({\bar \partial}\ln\varphi_\pm)=-\frac{1}{4}m^2R^2_5,
\end{eqnarray}
where use is made of eq.~(\ref{sum}) and 
\begin{eqnarray}
f(z)\equiv \frac{1}{4\pi}\bigg(\ln\vartheta_1\bigg(\frac{z}{2\pi}|\tau\bigg)
+\frac{1}{8\pi\tau_2}z^2\bigg).
\end{eqnarray}
For $m^2\neq 0$, we find the wave functions 
of massive modes outside the branes as follows
\begin{eqnarray}
{\tilde\phi}_{\pm}&=&{\tilde\phi}_{0,\pm} \,e^{c_\pm z-{\tilde c}_\pm{\bar z}}
\, \exp(\mp 2gq\sum_I\xi_I\int d^2 z' f(z-z')\Delta^2(z'-z_I))\label{massive}
\end{eqnarray}
where ${\tilde \phi}_{0,\pm}$ are overall constants to be fixed by
matching with the solutions inside the brane, 
and the integration constants $c_\pm$ and ${\tilde c}_\pm$ are related 
to each other by
\begin{eqnarray}
{\tilde c}_\pm=\frac{m^2R^2_5}{4}\bigg(c_\pm\pm\frac{gq}{8\pi^2\tau_2}
\sum_I\xi_I{\bar z}_I\bigg)^{-1}.
\end{eqnarray}
Our solution (\ref{massive}) is valid only in one half of the torus. 
The solutions in other regions are then obtained by applying the ${\bf Z}_2$ 
reflection. 
Moreover, considering the periodicities of the bulk solutions 
on the torus for $z\rightarrow z+2\pi$ 
and $z\rightarrow z+2\pi\tau$, we find the mass spectrum 
\begin{eqnarray}
m^2=\frac{4}{R^2_5}\bigg|c_\pm
\pm\frac{gq}{8\pi^2\tau_2}\sum_I\xi_I{\bar z}_I\bigg|^2
=\frac{4}{R^2_5}|{\tilde c}_\pm|^2, \label{mass}
\end{eqnarray}
with 
\begin{eqnarray}
c_\pm=\frac{1}{2}\bigg(\frac{n'}{\tau_2}+in\bigg),
\end{eqnarray}
where $n$ and $n'$ are integers. As a result, the mass spectrum depends only
on the quadratic FI terms, neither the log FI terms nor the brane thickness
do affect the mass spectrum. 
The structure of the resulting mass spectrum is so different from the $d=5$ case 
in the sense that there also appear linear terms in $\xi_I$'s. 
In particular, for $\sum_I\xi_Iz_I=0$, 
which is the case with no net dipole moments
coming from FI terms, even the nonzero localized FI terms do not modify 
the mass spectrum at all.
Even for $\sum_I\xi_Iz_I\neq 0$ with large FI terms,
there generically appears a normal KK tower of massive modes starting with large integers $n$ and $n'$ which cancel the shifts due to local FI terms.

Now let us consider as illustration 
the following nontrivial configuration of FI tadpoles: 
$\xi_1/3=-\xi_2=-\xi_3=-\xi_4$. 
This configuration can be obtained simply by considering 
a four-dimensional $U(1)$ anomaly free combination 
of {\it brane fields} with charges $q=+1$ for 
two fields at $z=0$, $q=-1$ for one field 
at each of the remaining fixed points and {\it one bulk field} with charge $q=+1$. 
Then, the mass spectrum becomes
\begin{eqnarray}
m^2=\frac{1}{R^2_5}\bigg(n\pm \frac{gq\xi_1}{6\pi}\bigg)^2
+\frac{1}{R^2_6}\bigg(n'\mp \frac{gq\xi_1}{6\pi}\bigg)^2.\label{spectrum}
\end{eqnarray}
This is equivalent to the mass spectrum without FI terms but with a constant Wilson line, 
$\langle (A_5-iA_6)\rangle=a_5/R_5-ia_6/R_6$ where 
$a_5=\pm gq\xi_1/(6\pi)=-a_6$. Moreover, since $q\xi_1>0$ for 
the bulk charged field with $q=+1$, there appears a simultaneous localization
of the bulk field at three fixed points other than $z=0$. 
For this case, the form of the zero-mode wave function is shown in Fig.~2. 

\begin{figure}[ht]
\begin{center}
\begin{minipage}{15.5cm}
\begin{minipage}[b]{8.25cm}
\begin{flushleft}
 \epsfig{file=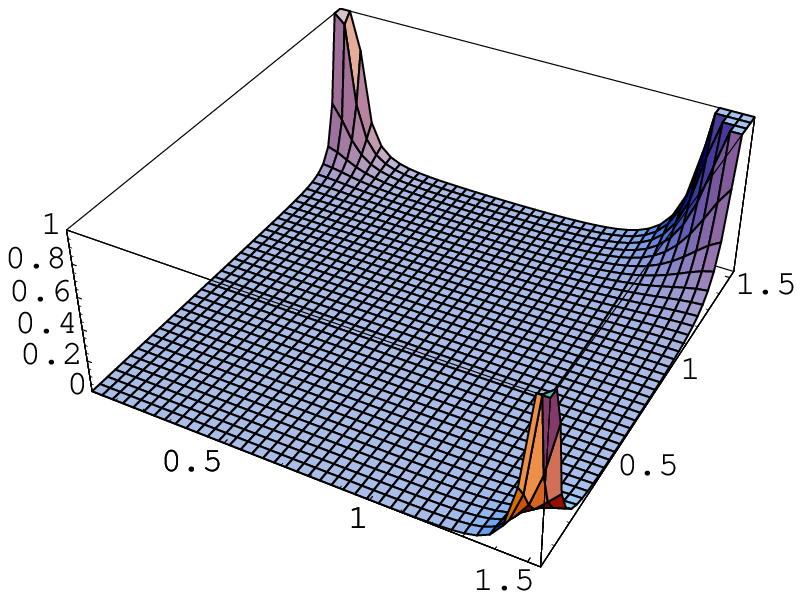,width=8.25cm} \\
\end{flushleft}
\par
\end{minipage}\hfill
\begin{minipage}[b]{6.75cm}
\begin{flushright}
 \epsfig{file=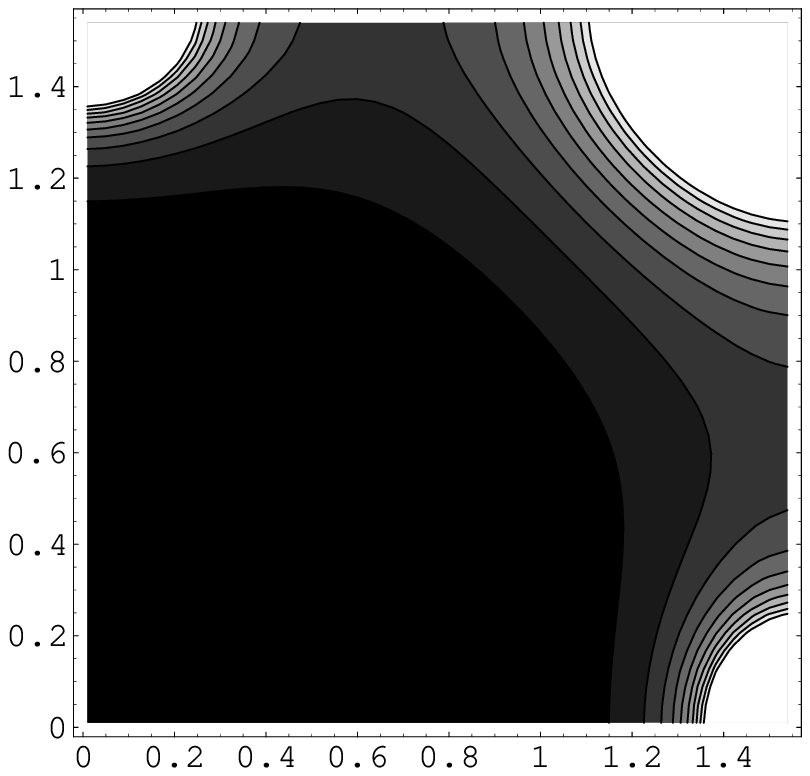,width=6.75cm} \\
\end{flushright}
\par
\end{minipage}
\end{minipage}
\caption{  \label{local}
Left figure: The wave function of the bulk zero-mode. It is drawn 
for one quarter of the torus
with a height rescaling by $10^{-3}$ and a chosen 
value of $g|q\xi_1|/(2\pi)=3$. 
Right figure: The contour presentation of the bulk zero-mode.}
\end{center}
\end{figure}
 
Comparing this mass spectrum with the one obtained in the
five-dimensional case (see formula (49) of 
ref. \cite{GrootNibbelink:2002qp}) we observe a qualitative difference.
There, the Kaluza-Klein excitations of the bulk mode became very
heavy with the cut-off $\Lambda$ and in the limit
$\Lambda\rightarrow\infty$ we just retained a massless zero mode
localized at a fixed point. Effectively the bulk field underwent
a dimensional transmutation and became a brane field. In the present 
six-dimensional case such a radical effect does not happen. The zero
mode bulk field shows a localization behaviour (as illustrated in 
Fig.~2), but the Kaluza-Klein excitations are not removed and the
bulk field retains its six-dimensional nature.

\section{Green-Schwarz mechanism for anomaly cancellation on the orbifold}

Localized FI-tadpoles will in general imply the presence of localized anomalies.
In this section, we consider the general case of 
local abelian and nonabelian anomalies coming from
bulk as well as brane fermions. 
In orbifolds, consistency requires the consideration of bulk 
anomalies\footnote{For the six-dimensional case, see Ref.~\cite{Erler}.}
as well as local anomalies appearing at the fixed points 
\cite{Arkani-Hamed, Serone, Pilo, Barbieri, serone,KKL, Asaka, LEE, Quiros}. 
The zero mode of a bulk fermion contributes to the local anomalies with equal
distribution at the fixed points and a brane fermion also leads to local 
anomalies at its fixed point. 
If eq.~(\ref{sum}) is fulfilled, we see 
that the $U(1)$ mixed gravitational anomalies are globally vanishing
for the supersymmetric vacuum. On the other hand, in general, 
it is not guaranteed that the $U(1)$ mixed gauge anomalies are also globally 
cancelled. 
In this section, we first give a general review on the anomaly cancellation 
on $d=6$ orbifolds from the field theory point of view.
Then, we show how a generalized 
Green-Schwarz (GS) mechanism \cite{Green} with various antisymmetric tensor
fields can lead to a cancellation of both bulk and local anomalies. 
In view of applications towards the heterotic string we explicitely
work out the conditions on the model where only one two-form field
strength cancels all anomalies.

The discussion in this paper is, of course, not restricted to 
the consideration 
of the pure six-dimensional case, but should also be understood as a
building block towards the ten-dimensional picture. It might thus occur
that there are more sectors of different dimensions which should be put
together to obtain the full model. In such a case formulae like 
(\ref{sum}) might have to be fulfilled only globally and not separately
for each sector. For the sake of simplicity we here restrict our
discussion to the six-dimensional case and deduce the conditions
for anomaly cancellation in that scheme. The generalization to more
complex systems should then be straightforward.

So let us first consider the bulk anomaly.
The factorizable $d=6$ bulk anomaly comes from two different anomaly 
polynomials
\begin{eqnarray}
I^{(1)}_8&=&X_4{\tilde X}_4, \\
I^{(2)}_8&=&X_2 X_6,
\end{eqnarray}
where all forms on the right-hand sides are Casimir invariants 
containing the gravitational ($R$) and/or gauge field strengths ($F$).
The first one contains only second order Casimir invariants 
while the second one involves cubic anomalies with $U(1)$.  
In principle, 
the bulk anomalies of the types $I^{(1)}_8$ and $I^{(2)}_8$ can be cancelled 
by a $d=6$ GS term with a bulk 2-form and a bulk 4-form, 
respectively. However, 
the existence of the $I^{(2)}_8$ anomaly renders the $U(1)$ massive due to the nonzero
six-dimensional dual axion coupling\footnote{This is the analogue of 
the cancellation of the $d=4$ local anomalies with the brane 
2-forms \cite{serone,Antoniadis:2002cs}.}  with the $U(1)$, 
irrespective of the absence of the four-dimensional anomaly.

On the orbifold, there also appear local anomalies at the fixed points 
which take the forms as follows 
\begin{eqnarray}
I^{(1)}_6&=&Y_2Y_4, \\
I^{(2)}_6&=&Y_6, 
\end{eqnarray}
where $Y_2$ denotes the field strength of the bulk $U(1)$ 
and $Y_4,Y_6$ are other Casimir invariants. 
Thus the local abelian and non-abelian anomalies somehow modify 
$I^{(1)}_8$ and $I^{(2)}_8$, respectively. 

It has been already shown that the local reducible $U(1)$ 
anomalies of type $I^{(1)}_6$ can be cancelled
by the GS mechanism with brane 2-forms \cite{serone,Antoniadis:2002cs} 
or a bulk two-form \cite{GrootNibbelink:2003gb,Quiros}, irrespectively of
whether they are globally vanishing or not. 
For the $U(1)$ anomaly of type $I_6=\sum_I c_I FY_4|_{z_I}$, with $c_I$
being arbitrary numbers, 
the relevant GS Lagrangian with four brane axions becomes
\begin{eqnarray}
S_{GS}=\int \sum_I\bigg[-\frac{1}{2}|da^I-A|^2-c_I a_I Y_4\bigg]
\delta^2(z-z_I)d^2 z
\end{eqnarray}
where $F=dA$ is the $U(1)$ field strength 
and the axion $a_I$ at each fixed point transforms 
as $\delta a_I=\Lambda(z_I)$ under the gauge transformation, 
$\delta A=d\Lambda$. In this case, independently of whether 
$\sum_I c_I=0$ or not, it can be seen that nonzero mass terms for $U(1)$ appear
due to the local axion couplings \cite{serone,Antoniadis:2002cs}. 

On the other hand, the local irreducible  
anomalies of type $I^{(2)}_6$, which contain the local $U(1)^3$ 
and non-abelian anomalies, can be cancelled 
by the GS mechanism with a bulk four-form \cite{serone,Quiros}. 
For the $U(1)$ anomaly of type $I_6=\sum_I c'_I F^3|_{z_I}$, with $c'_I$
being numbers satisfying $\sum_I c'_I=0$,
the relevant GS Lagrangian with one bulk four-form becomes
\begin{eqnarray}
S_{GS}&=&\int -\frac{1}{2}|dC_4-Y_5|^2
-\sum_I c'_I C_4\delta^2(z-z_I)d^2z \nonumber \\
&=&\int \eta Y_5
\end{eqnarray}
where $C_4$ is integrated out in the last line 
and the one-form $\eta$ is defined as
\begin{eqnarray}
d\eta=\sum_I c'_I\delta^2(z-z_I).
\end{eqnarray}
Likewise, note that $Y_6=d F^3$, and $\delta C_4=Y^1_4$ 
with $\delta Y_5=d Y^1_4$ under the gauge transformation. 
In this case, the GS mechanism involving
the bulk four-form field is valid only for {\it globally vanishing} local
anomalies because the bulk four-form appears
only as massive KK modes in the four-dimensional 
effective field theory \cite{serone,Quiros}. 
Therefore, if local $U(1)^3$ anomalies are not reducible,
there should be no integrated $U(1)^3$ anomaly in our model. 

In the case of orbifold compactification of the heterotic string, however, 
the possible matter content should be much more restricted 
because there is only one two-form available for the GS mechanism.
For instance, there would not appear either local irreducible 
anomalies of type $I^{(2)}_6$ or bulk anomalies 
of type $I^{(2)}_8$ \cite{Gmeiner:2002es,GrootNibbelink:2003gb}. 
Now let us consider the simultaneous cancellation of bulk and local anomalies 
by the GS mechanism with only one bulk 2-form in $d=6$. 
The total anomaly polynomial on $T^2/{\bf Z}_2$ we are considering is 
\begin{eqnarray}
i(2\pi)^3 I_8=\frac{1}{2}X_4{\tilde X}_4
+i(2\pi)^3 \sum_I (\frac{1}{4}I_6|_{r_B}+I_6|_{r_I}) \delta^2(z-z_I)d^2 z
\end{eqnarray}
where 
\begin{eqnarray}
X_4&=&{\rm tr}R^2-\sum_A \alpha^{(1)}_A {\rm tr}_F F^2_A, \\
{\tilde X}_4&=&\frac{1}{16}({\rm tr}R^2
-\sum_A \alpha^{(2)}_A {\rm tr}_F F^2_A),
\end{eqnarray}
and $I_6|_{r}$ denotes
the four-dimensional anomaly polynomial for
the bulk (brane) fermion in the representation $r=r_{B(I)}$ as 
\begin{eqnarray}
i(2\pi)^3 I_6|_{r}
=-\frac{1}{48}{\rm tr}_{r}F {\rm tr}R^2
+\frac{1}{6}{\rm tr}_{r}F^3+\frac{1}{2}{\rm tr}_{r}(F F^2_A). 
\label{4danomaly}
\end{eqnarray}
Here ${\rm tr}_F$ means the trace for the fundamental representation
and $\alpha^{(1)}_A$ is given as $\alpha^{(1)}_A=\frac{1}{l_A(F)}$
for groups realized at level 1 Kac-Moody algebras \cite{Erler}
where $l_A(F)={\rm tr}_F(T_AT_A)$ is the index of the fundamental
representation of group $A$, and $\alpha^{(2)}_A$ depend on the matter
representations \cite{Erler}.
Moreover, we note that the factor $\frac{1}{2}$ in the bulk anomaly 
comes from the reduction of the fundamental region
on $T^2/{\bf Z}_2$ and the factor $\frac{1}{4}$ in the local anomaly comes from
the number of fixed points. $F$ and $F_A$ are the abelian and 
non-abelian gauge field strengths. 

In order for the Green-Schwarz mechanism 
with one bulk two-form to work for the 
anomaly cancellation, we need the universal relation between abelian anomalies 
at each fixed point as
\begin{eqnarray}
\frac{1}{48}{\rm tr}Q_I=\frac{1}{6}{\rm tr}Q^3_I=\frac{1}{2}{\rm tr}(l_A(r)Q_I),
\end{eqnarray}
where ${\rm tr}Q_I=\frac{1}{4}{\rm tr}(q)+{\rm tr}(q_I)$ and 
$l_A(r)$ is the index of representation $r$ under the gauge group $A$.
In this case, other global abelian anomalies are also vanishing due to 
$\sum_I {\rm tr}Q_I=0$ from eq.~(\ref{sum}). 
That is to say, the $d=4$ anomaly polynomial (\ref{4danomaly}) becomes reducible 
at each fixed point as
\begin{eqnarray}
I_6|_{r}=\alpha (FX_4)|_r,
\end{eqnarray}
with $\alpha=\frac{i}{(2\pi)^3}\frac{1}{48}$.
Then, the total $d=6$ anomalies are given by
\begin{eqnarray}
{\cal A}_6=\int I^1_6+\sum_I I^1_4|_I\delta^2(z-z_I)d^2 z, \label{anomaly}
\end{eqnarray}
where $I^1_6$ and $I^1_4$ are descendents of the bulk and local anomaly
polynomials, respectively. 

The Green-Schwarz Lagrangian with one two-form is  
\begin{eqnarray}
S_{GS}&=&\int -\frac{1}{2}*(dB-X_3)(dB-X_3)\nonumber \\
&-&\bigg(\frac{-i}{2(2\pi)^3}{\tilde X}_3 
+\sum_I \alpha (\frac{1}{4}A|_{r_B}+A|_{r_I})
\delta^2(z-z_I)d^2 z\bigg)dB \nonumber \\
&+&\bigg(\beta {\tilde X}_3+\sum_I\beta_I (\frac{1}{4}A|_{r_B}+A|_{r_I})
\delta^2(z-z_I)d^2 z\bigg)X_3.\label{gsterm}
\end{eqnarray}
Note that $X_3$ and ${\tilde X}_3$ are defined 
from $X_4=dX_3$ and ${\tilde X}_4=d{\tilde X}_3$, and $\delta X_3=dX^1_2$
and $\delta{\tilde X}_3=d{\tilde X}^1_2$ under the gauge transformation.
Thus, the modified kinetic term in eq.~(\ref{gsterm}) 
requires that $\delta B=X^1_2$ under the gauge transformation. 
Therefore, we find that the bulk and local anomalies
in eq.~(\ref{anomaly}) are cancelled by the variation of the above 
Green-Schwarz
action with $\beta=\frac{-i}{4(2\pi)^3}$ and $\beta_I=\frac{1}{3}\alpha$.

Before closing this section, 
let us make a remark on the axion gauge coupling at the fixed points. From 
the second line in eq.~(\ref{gsterm}), we find that the two-form field
generically has different couplings with the $U(1)$ gauge 
potentials at different fixed 
points:
\begin{eqnarray}
&&-\int\sum_I\alpha (\frac{1}{4}A|_{r_B}+A|_{r_I})\delta^2(z-z_I)d^2 z dB 
\nonumber \\
&=&-\sum_I \alpha(\frac{1}{4}{\rm tr}(q)+{\rm tr}(q_I))\int d^4 x 
A_a(x,z_I)\partial^a b(x,z_I)\label{axion}
\end{eqnarray}
where under the local duality transformation of the two-form, 
$b(x,z_I)$ is considered as a brane projection of the ${\bf Z}_2$ 
even $B_{56}$ as the following
\begin{eqnarray}
\partial_{[a}B_{bc]}(x,z_I)
&=&\varepsilon_{abcd 5 6}\,\partial^{[d} B^{56]}(x,z_I)\nonumber \\
&\equiv& \varepsilon_{abcd}\,\partial^d b(x,z_I).
\end{eqnarray} 
Therefore, plugging into eq.~(\ref{axion}) 
the zero modes of $A_a(x,z)$ and $b(x,z)$ which are constant in the bulk, 
the effective axion coupling becomes vanishing 
due to eq.~(\ref{sum}) after the sum of the local axion gauge couplings. 
Consequently, we find that the local axion gauge couplings incorporated 
for the local anomaly cancellation do not break the $U(1)$.

\section{Concluding remarks}

The consideration of higher dimensional brane world schemes leads to
fascinating possibilities to extend the standard model of particle
physics. The most promising starting point is provided by superstring
theory in $d=10$ or M-theory in $d=11$, with six or seven compactified
dimensions, respectively. In the framework of $d=10$ supersymmetric 
string theories, extra space dimensions typically appear in complex 
pairs. Such theories contain matter fields on branes of various 
dimensionalities. The profile of the higher dimensional bulk fields
in the presence of (lower dimensional) brane fields is of particular
importance for the phenomenological properties of a given scheme.
Earlier studies in the co-dimension 1 case revealed a specific localization
phenomenon in the presence of localized FI-tadpoles, where a ($d=5$) bulk
field dimensionally transmuted to a ($d=4$) brane field. In this paper we 
examined the more complicated case of co-dimension 2 in the framework
of a supersymmetric orbifold theory in 6 space-time dimensions.
Because of the holomorphic structure of the one complex (= two real)
extra dimensions we were able to compute the 
Kaluza-Klein mass spectrum and the
wave function of the zero-mode bulk field explicitly. The key ansatz
is given in equation (\ref{W}), where the holomorphic structure is transparent.
Again we find a localization phenomenon of the bulk zero mode
(see equation (\ref{final}) and for illustration figure 2), but the situation
differs from the co-dimension 1 case, as the bulk field retains its 
six-dimensional nature. The spectrum of massive modes equation (\ref{spectrum})
is equivalent to a spectrum in the presence of a constant Wilson line.

The co-dimension 2 case should be relevant for the discussion
of those compactified superstring theories in $d=10$, where we deal 
in general with 3 complex extra dimensions and the presence of
3-branes ($d=4$) and 5-branes ($d=6$). In addition the co-dimension 1
case could find its application in $d=11$ M-theory and/or string
theories that contain 3-branes and 6-branes of $d=7$. The potential 
problem of localized gauge or gravitational anomalies is cured
with the help of  a generalized Green-Schwarz mechanism.

\section*{Acknowledgments}
This work is supported by the
European Community's Human Potential Programme under contracts
HPRN-CT-2000-00131 Quantum Spacetime, HPRN-CT-2000-00148 Physics Across the
Present Energy Frontier and HPRN-CT-2000-00152 Supersymmetry and the Early
Universe. HML was supported by priority grant 1096 of the Deutsche
Forschungsgemeinschaft.

\begin{appendix}
\section{Notations and Conventions\label{appa}}

Our conventions are six-dimensional generalizations of the ones used in \cite{Zucker:1999ej, Zucker:2003qv}.
The metric is $\eta_{AB}=diag(+-----)$; $A,B=0,1,2,3,5,6$ are six-dimensional indices and $a, b=0,\ldots, 3$ are four-dimensional ones.

Our gamma-matrices are antisymmetrized with strength one. As always in even dimensional spacetimes, one can introduce a 
chirality operator which is defined in the six-dimensional case by (with $\varepsilon^{012356}=+1$)
\begin{equation}
\Gamma^7=\frac{1}{6!}\varepsilon^{A_1\ldots A_6}\Gamma_{A_1\ldots A_6}.
\end{equation}


An explicit representation for the gamma-matrices is
\begin{equation}
\Gamma^a=\tau^1\otimes\gamma^a,\qquad \Gamma^5=\tau^1\otimes \gamma^5,\qquad \Gamma^6=-i\tau^2\otimes \unity_4,\label{rep1}
\end{equation}
where $\gamma^a$ and $\gamma^5$ are the standard four-dimensional gamma matrices, with
\begin{equation}
\gamma^5=-\gamma^0\gamma^1\gamma^2\gamma^3.
\end{equation}
In this basis, the six-dimensional chirality operator is diagonal:
\begin{equation}
\Gamma^7=-\tau^3\otimes \unity_4.
\end{equation}
The charge conjugation is then
\begin{equation}
C=-i\tau^2\otimes {\cal C},\label{rep4}
\end{equation}
where $\cal C$ is the five-dimensional charge conjugation.




%
Spinors carry an $SU(2)_{\cal R}$ index $i=1,2$. 
One can impose a symplectic Majorana condition
\begin{equation}
\bar\psi_i\equiv(\psi^i)^\dagger\Gamma^0=\varepsilon_{ij}(\psi^j)^TC,\label{SMB}
\end{equation}
where the charge conjugation matrix $C$ fulfills the following relations:
\begin{equation}
\Gamma_A^T=-C\Gamma_A C^{-1},\qquad C=C^T,\qquad C^\dagger C=1.
\end{equation}

Whenever possible we suppress the $SU(2)_{\cal R}$ indices. In this case the summation convention is from southwest to northeast, for example
\begin{equation}
\bar{\psi}\Gamma^A\lambda\equiv\bar{\psi}_i\Gamma^A\lambda^i,\qquad \bar{\psi}\vec{\tau}\Gamma^A\lambda\equiv\bar{\psi}_i\Gamma^A\lambda^j(\vec{\tau})^i{}_j,\label{conv1}
\end{equation}
with $(\vec{\tau})^i{}_j$ the standard Pauli matrices and the arrows denotes generically the vector representation of $SU(2)_{\cal R}$.

A fundamental and rather useful identity is
\begin{equation}
\bar\psi_i\Gamma^{A_1}\ldots\Gamma^{A_n}\lambda^j=(-1)^n\varepsilon_{ik}\varepsilon^{jl}\bar{\lambda}_l\Gamma^{A_n}\ldots\Gamma^{A_1}\psi^k,
\end{equation}
from which one easily deduces the symmetry properties of fermionic bilinears.
%
%

For four spinors $\psi,\lambda,\chi,\xi$ there are two Fierz identities possible, depending on the relative chirality of the fields:

If $\psi$ and $\xi$ have the same chirality, we have
\begin{eqnarray}
(\bar{\chi}\psi)(\bar{\lambda}\xi)&=&
-\frac{1}{8}(\bar{\chi}\xi)(\bar{\lambda}\psi)
-\frac{1}{8}(\bar{\chi}\vec{\tau}\xi)(\bar{\lambda}\vec{\tau}\psi)\nonumber \\
&+&\frac{1}{16}(\bar{\chi}\Gamma_{AB}\xi)(\bar{\lambda}\Gamma^{AB}\psi)
+ \frac{1}{16}(\bar{\chi}\vec{\tau}\Gamma_{AB}\xi)
(\bar{\lambda}\vec{\tau}\Gamma^{AB}\psi).
\end{eqnarray}

If $\psi$ and $\xi$ have opposite chirality, we have
\begin{eqnarray}
(\bar{\chi}\psi)(\bar{\lambda}\xi)&=& 
-\frac{1}{8}(\bar{\chi}\Gamma_A\xi)(\bar{\lambda}\Gamma^A\psi)
-\frac{1}{8}(\bar{\chi}\vec{\tau}\Gamma_A\xi)
(\bar{\lambda}\vec{\tau}\Gamma^A\psi)\nonumber \\
&+&\frac{1}{48}(\bar{\chi}\Gamma_{ABC}\xi)(\bar{\lambda}\Gamma^{ABC}\psi)
+\frac{1}{48}(\bar{\chi}\vec{\tau}\Gamma_{ABC}\xi)
(\bar{\lambda}\vec{\tau}\Gamma^{ABC}\psi).
\end{eqnarray}

\end{appendix}


\end{document}